\documentclass[twocolumn,showpacs,preprintnumbers,amsmath,amssymb,floatfix,nofootinbib]{revtex4-1}
\usepackage{amsmath,amsfonts,bbm,euscript,hyperref}
\usepackage{graphicx}
\usepackage{dcolumn}
\usepackage{bm}
\usepackage{epsfig}
\epsfclipon

\newcommand{\nco}{\newcommand}
\nco{\beq}{\begin{equation}} \nco{\eeq}{\end{equation}}
\nco{\beqa}{\begin{eqnarray}} \nco{\eeqa}{\end{eqnarray}}
\def\be{\begin{equation}}
\def\ee{\end{equation}}
\def\baray{\begin{eqnarray}}
\def\earay{\end{eqnarray}}

\nco{\lra}{\leftrightarrow}

\nco{\sss}{\scriptscriptstyle} \nco{\dphi}{\varphi}
\nco{\lsim}{\mbox{\raisebox{-.6ex}{~$\stackrel{<}{\sim}$~}}}
\nco{\gsim}{\mbox{\raisebox{-.6ex}{~$\stackrel{>}{\sim}$~}}}

\def\IK{\relax{\rm I\kern-.20em K}}
\def\IM{\relax{\rm I\kern-.20em M}}

\def\lsim{\mbox{\raisebox{-.6ex}{~$\stackrel{<}{\sim}$~}}}
\def\gsim{\mbox{\raisebox{-.6ex}{~$\stackrel{>}{\sim}$~}}}
\def\sss{\scriptscriptstyle}

\begin{document}

\preprint{UMN-TH-3416/15, FTPI 15/1}

\title{Affleck-Dine Sneutrino Inflation}

\author{Jason L. Evans$^{1,2}$, Tony Gherghetta$^1$, Marco Peloso$^1$}

\affiliation{
\centerline{$^1$School of Physics and Astronomy, University of Minnesota, Minneapolis, MN 55455, USA}
\centerline{$^2$William I. Fine Theoretical Physics Institute, University of Minnesota, Minneapolis, MN 55455, USA}
Email: {\tt jlevans@umn.edu, tgher@physics.umn.edu, peloso@physics.umn.edu}}

\begin{abstract}
Motivated by the coincidence between the Hubble scale during inflation and the typical see-saw neutrino mass  scale, we present  a supergravity model where the inflaton is identified with a linear combination of right-handed sneutrino fields. The model accommodates an inflaton potential that is flatter than quadratic chaotic inflation, resulting in a measurable but not yet ruled out tensor-to-scalar ratio. Small CP-violation in the neutrino mass matrix and supersymmetry breaking 
yield an evolution in the complex plane for the sneutrino fields. This induces a net lepton charge that, via the  Affleck-Dine mechanism,  can be the origin of the observed baryon asymmetry of the universe.
\end{abstract}


\maketitle

\section{Introduction}
\vspace{-4mm}

Inflation is a postulated period of accelerated expansion of the universe that preceded  the  hot-big bang era and solves a number of cosmological problems \cite{Linde:2005ht}. The simplest models of inflation predict a nearly scale-invariant spectrum for the scalar density perturbations $P_{\rm s}$ and tensor gravitational waves, $P_t$. Density perturbations imprint the Cosmic Microwave Background (CMB) with temperature anisotropies and polarization, and seed the large scale structures of the universe. Both the large scale structure and, particularly, the CMB are in excellent agreement with the predictions from inflation \cite{Ade:2013uln}.
On the other hand, there is not yet compelling evidence for gravitational waves, and the current upper bound on the tensor-to-scalar ratio is $r \equiv \frac{P_t}{P_s} \lsim 0.11$ (the precise limit depends on the priors and on the data set used \cite{Ade:2013uln}). This gravitational wave signal may be detected in the near future if $r \gsim 0.01-0.05$ \cite{Baumann:2008aq}. In conventional models, this requires that the inflaton $\phi$
spans a range $\Delta \phi \gsim {\cal O } \left( M_p \right)$ during the stage of inflation in which the CMB modes were generated \cite{Lyth:1996im}. Here $M_p \simeq 2.48 \times 10^{18} \, {\rm GeV}$ is the (reduced) Planck mass.

This class of models, called ``large-field inflation'', is particularly challenging to obtain in particle physics, because quantum corrections to the inflaton potential $\Delta V = \frac{\phi^{4+n}}{M^n}$ will in general spoil the required flatness of the potential, unless $M \gg M_p$ \cite{Weinberg:2008hq}. In supergravity, it is hard to achieve flatness even at the classical level,  due to the exponential dependence on the  K\"ahler potential $K$ in the scalar potential $V$,
\begin{equation}
V = {\rm e}^{\frac{K}{M_p^2}} \left[ K^{i{\bar j}} \left( D_i W \right) \left( D_j W \right)^\dagger - \frac{3 \vert W \vert^2}{M_p^2} \right] \,,
\label{V}
\end{equation}
where $W$ is the superpotential,  $D_i \equiv \partial_i W + W \partial_i K$, and the subscripts $i,j$ are scalar field labels. A possible solution to this problem is to protect the flatness of $V$ with a shift symmetry; this is true both in the general case, \cite{Freese:1990rb} - see \cite{Pajer:2013fsa} for a review - and in supergravity  \cite{Kawasaki:2000yn,Kallosh:2010ug,Kallosh:2011qk}, where the symmetry prevents a field from entering in $K$.

A second challenge for model building is to embed inflation into a more complete framework of particle physics.  An excellent possibility, first considered in \cite{Murayama:1993xu,Murayama:2014saa}, is to identify the inflaton with one of the right-handed sneutrino fields in the supersymmetric standard model. Assuming for simplicity just two right-handed neutrino superfields $N_{1,2}$, with complex scalar field components $\phi_{1,2}$, we consider the following Kahler and superpotential 
\begin{eqnarray}
\label{kahler}
K &=& \frac{1}{2} \vert N_1 + N_1^\dagger \vert^2 + \vert N_2 \vert^2\,,\\
W &=& \frac{1}{2} m_1 N_1^2+m N_1 N_2 + \frac{1}{2} m_2 N_2^2+ m_{3/2} M_p^2\;,
\label{model}
\end{eqnarray}
where $m, m_{1,2}$ are supersymmetric masses, and the gravitino mass  $m_{3/2}$ is a supersymmetry-breaking parameter~\footnote{We actually need to fix the minimum of the scalar potential to zero (this corresponds to the standard tuning of the cosmological constant to zero), through a cancellation  between the F-terms of the supersymmetry breaking term and the superpotential. Assuming that the supersymmetry breaking $F-$terms do not depend on the neutrino superfields amounts to introducing an extra term $\Delta V = 3 m_{3/2}^2 M_p^2 {\rm e}^{K/M_p^2}$  in (\ref{V}).}. A special feature of  (\ref{kahler}) is that, due to a shift symmetry on $N_1$, the imaginary part of  $\phi_1$ is absent from $K$ and therefore can be  (predominantly) identified with the inflaton.  In the model of  \cite{Murayama:2014saa}, a specific mass texture $m_1=m_2=0$ is assumed, resulting in $\phi_2=0$ during inflation, and a chaotic inflationary potential $V = \frac{1}{2} m^2 \phi_1^2$ (plus negligible corrections). In addition the third right-handed neutrino field is assumed to be lighter and thermally produced after inflation, generating a net lepton asymmetry when it decays. This is subsequently converted to a baryon asymmetry by sphalerons via thermal leptogenesis~\cite{Fukugita:1990gb}.

Instead in this Letter we consider a more generic mass texture in the superpotential (\ref{model}) (which could be further generalized to three fields) and show that a nonzero $m_1$ gives rise to $\phi_2\neq 0$ during inflation. For real $m_1$, the inflationary evolution occurs in the imaginary $\phi_{1,2}$ plane. Interestingly the inflationary potential along this trajectory is flatter than the simple quadratic potential $V = \frac{1}{2} m^2 \phi_1^2$, leading to a detectable but smaller value of $r$ compared to massive chaotic inflation.

Furthermore, while $m$ can always be made real by changing the phase of $N_2$, no corresponding change of phase is allowed  for $N_1$, due to the structure of $K$.  Therefore $m_1$ is in general complex.
A small imaginary part of $m_1$ does not substantially modify the inflationary results obtained for real $m_1$, but allows for a CP-violating and lepton number, $n_L$, carrying trajectory of the right-handed sneutrino fields where:
\begin{eqnarray}
n_L &\equiv& i \left[ \phi_1^*\partial_0 \phi_1 + \phi_2^* \partial_0 \phi_2 \right] + h.c.
\label{nL}
\end{eqnarray}
When the sneutrino decays, $n_L$ is transferred to the decay products. This is a realisation of the Affleck-Dine mechanism \cite{Affleck:1984fy}  in which the baryon asymmetry of the universe originates from the evolution of a scalar condensate.

\vspace{-4mm}
\section{Inflation}
\label{sec:inflation}
\vspace{-4mm}

To compute the scalar field potential following from  (\ref{model}), we define the real and imaginary components of $\phi_i$ as $\phi_i \equiv \frac{1}{\sqrt{2}} (\phi_{iR} + i \phi_{iI})$ and introduce dimensionless parameters through the rescalings $m_1 \equiv m \left( \mu_R + i \, \mu_I \right)$ and $m_{3/2} \equiv m \, \mu_{3/2}$. Furthermore, we neglect $m_2$ and assume $\mu_I ,\, \mu_{3/2}  \ll \mu_R \ll 1$ so that $\mu_I$ and $\mu_{3/2}$ can be disregarded during inflation. In the limit $\mu_I = \mu_{3/2} =0$, the model admits the stable solutions  $\phi_{1R} = \phi_{2R}=0$ and the potential for the nonvanishing directions reads
\begin{eqnarray}
&& \!\!\!\!\!\!\!\! V  = \frac{1}{2}m^2 \phi_{1I}^2 \,  {\rm e}^{\frac{\phi_{2I}^2}{2 M_p^2}} \,
\Bigg\{ \left( 1 + \mu_R^2 \right) - \frac{3}{8} \gamma^2 - \left( 1 - \frac{2 M_p^2}{\phi_{1I}^2} \right) \frac{\phi_{2I}}{M_p} \nonumber\\
&& -  \frac{1}{2} \left[ \left( 1 -  \frac{2 M_p^2}{\phi_{1I}^2} \right) - \frac{\gamma^2}{8} \right] \frac{\phi_{2I}^2}{M_p^2} + \frac{\gamma}{4} \frac{\phi_{2I}^3}{M_p^3} + \frac{1}{4} \frac{\phi_{2I}^4}{M_p^4} \Bigg\} \,,
\label{V-inf}
\end{eqnarray}
where $\gamma \equiv \mu_R \, \frac{\phi_{1I}}{M_p}$. The field $\phi_{2I}$ is sufficiently massive during inflation (see below), and can be integrated out. Namely, we compute the equation of motion for $\phi_{2I}$ following from (\ref{V-inf}), and we write the solution as $\phi_{2I} \left[ \phi_{1I} \right]$. Inserting this solution back into (\ref{V-inf}), we obtain the single-field effective inflationary model $V_{\rm eff} \left( \phi_{1I} \right) = V \left( \phi_{1I}, \phi_{2I} \left[ \phi_{1I} \right] \right)$. In doing so, we can obtain a remarkably simple solution, if we disregard the second term in each of the three round parentheses in eq. (\ref{V-inf}). Thus, one can verify that $\frac{\partial V}{\partial \phi_{2I}} =0$ is solved by $\gamma = \frac{2 \phi_{2I}^3}{4 M_p^3 - \phi_{2I}^2 \, M_p}$. This relation can then be inverted to give
\begin{equation}
\frac{\phi_{2I}}{M_p} \simeq 2^{1/3} \, \gamma^{1/3} - \frac{\gamma}{6} + {\cal O } \left( \gamma^{5/3} \right) \,,
\label{XI-gamma}
\end{equation}
which, once substituted into (\ref{V-inf}), gives the effective inflationary potential
\begin{equation}
V_{\rm eff} \simeq \frac{1}{2}m^2 \phi_{1I}^2 \left[ 1 - \frac{3}{2^{5/3}} \left( \frac{\mu_R \, \phi_{1I}}{M_p} \right)^{4/3}
 - \frac{13}{24} \, \left(  \frac{\mu_R \, \phi_{1I}}{M_p} \right)^2 \right] \,.
\label{V-eff}
\end{equation}
The corrections in the square parenthesis arise from integrating out $\phi_{2I}$, and cause a flattening of the potential.~\footnote{A flattened inflation potential from integrating heavy fields out has also been obtained in \cite{Dong:2010in,McAllister:2014mpa,Buchmuller:2015oma}.} A flatter potential corresponds to a smaller  $r$. 

Excellent agreement is found when the evolution of $\phi_{1I}$ is computed using the effective
single-field potential (\ref{V-eff}) and compared to the evolution of the $\left\{ \phi_{1I} , \phi_{2I} \right\}$ pair in the full potential (\ref{V-inf}). The value of $r$ and the spectral tilt $n_s$ (defined from the scaling with wavenumber $k$ of the power spectrum of the density perturbations, $P_{\rm s} \propto k^{n_s-1}$) can then be computed numerically using the effective single-field potential (\ref{V-eff}).

Employing the standard procedure to compute  $r$ in slow-roll inflation (see for instance \cite{Martin:2014vha}) we obtain
\begin{equation}
r \simeq \frac{8}{N} \left[ 1 - \frac{7 \times 2^{2/3}}{5} \, N^{2/3} \, \mu_R^{4/3} - \frac{13}{8} \, \mu_R^2 \right] \, ,
\label{r}
\end{equation}
where $N$ is the number of e-folds of inflation. The  result $r=\frac{8}{N}$ of massive chaotic inflation is recovered at $\mu_R=0$.

\begin{figure}[h]
\includegraphics[width=0.4\textwidth,angle=0]{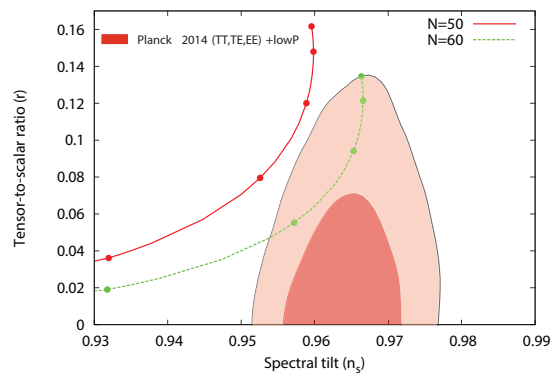}
\caption{Values of the spectral tilt $n_s$ and the tensor-to-scalar ratio $r$ for  $N=50$ (red-solid lines) and $N=60$ (green dotted line). Each line is obtained by varying $\mu_R$. From top to bottom, the points on the lines correspond to the values $\mu_{R} = \left( 0, 1, 2, 3, 4 \right) \times 10^{-2}$, respectively. The Planck contours are obtained from 
\cite{Finelli}. The top points $\mu_R=0$ correspond to quadratic chaotic inflation. 
}
\label{Fig:inflation}
\end{figure}

These results are shown in Figure \ref{Fig:inflation}. The numerical result for $r$ is in very good agreement with (\ref{r}). As is clear from the figure, a measurement of $r$ could determine the value of $\mu_R$ in the model. Moreover, the amplitude of $P_s$ determines the overall scale, $m$ of the potential~\cite{Martin:2014vha}. The value is only weakly sensitive to $\mu_R$ in the interval shown in Figure \ref{Fig:inflation}: for $\mu_R =0$, and $N=50 (60)$ we recover the result $m \simeq 1.9 (1.6) \times 10^{13} \, {\rm GeV}$ of massive chaotic inflation, while for $\mu_R=0.04$ we obtain
$m \simeq 1.5 (1.1) \times 10^{13} \, {\rm GeV}$. This coincides with the typical see-saw neutrino mass scale. 

Finally, let us justify the approximations used in  (\ref{V-eff}). For definiteness,  we set $\mu_R = 0.03$, (well within the experimental limit shown in the figure) and $N=60$ (the results for $N=50$ are very similar).
Evaluating numerically the potential (\ref{V-inf}) and its derivatives along the inflaton trajectory, we find the ratio between the mass of the $\phi_{2I}$ field ($m_{\phi_{2I}}$), and the Hubble rate\footnote{We recall that $H = \frac{\dot{a}}{a}$, where $a$ is the scale factor and dot denotes differentiation with respect to time $t$.}  (namely, $\frac{m_{\phi_{2I}}}{H} = \sqrt{\frac{3 M_p^2 V,_{\phi_{2I} \phi_{2I}}}{V}}$) to be $\simeq 3.6$. Therefore, it is consistent to integrate $\phi_{2I}$ out, and disregard its fluctuations. Secondly, we see that the terms that we disregarded in (\ref{V-inf}), provide at most  a $1.1\%$ correction.
This is insignificant compared to the effects obtained in our final expression (\ref{V-eff}) (namely, the last two terms in the square parenthesis amount to
$-0.26$ and $-0.08$, respectively). Thirdly, we note that integrating out the field $\phi_{2I}$ also provides a correction to the kinetic term of $\phi_{1I}$ in the single-field description (related to the fact that $\phi_{2I}$ contributes in part to the inflationary trajectory),  $E_{\rm kin} = \frac{1}{2} \left[ 1 + \left( \frac{ \partial \phi_{2I}}{\partial \phi_{1I}} \right)^2 \right] \dot{\phi}_{1I}^2$, where the field derivative is evaluated on the solution  $\phi_{2I} \left[ \phi_{1I} \right]$. The correction amounts to $3.2 \times 10^{-4}$ , and is therefore also negligible.

\section{Leptogenesis}
\label{sec:lepto}

One can verify that,  setting $\mu_I=0$ in the full potential  (\ref{V}), leads to
 $\phi_{1R}=\phi_{2R}=0$, while setting $\mu_{3/2} = 0$ leads to  $\phi_{1R} =0\,,\; \phi_{2R} = - \frac{\mu_I}{\mu_R} \phi_{2I}$. Both cases lead to $n_L = 0$ in eq. (\ref{nL}). Therefore the lepton number in this mechanism can be suppressed by the smallness of these parameters, $n_L = {\cal O } \left( \mu_I \, \mu_{3/2} \right)$. The exact evolution of the fields after inflation cannot be computed analytically. However, with some approximations we can  reproduce the correct order of magnitude of the baryon asymmetry. 

In the basis  $\Phi=\left\{ \phi_{1R},\phi_{2R},\phi_{1I},\phi_{2I} \right\}$  the scalar equations of motion  can be formally written as
\begin{equation}
\ddot{\Phi}_i + 3 H \dot{\Phi}_i + \partial_i V = 0 \,.
\label{eom}
\end{equation}
Using the fact that  $\vert \Phi_i \vert \ll M_p$ after inflation, we expand $V$ at quadratic order in the fields. Oscillations of massive fields effectively have the same equation of state as matter, and so~\footnote{This relation, as well as the approximation in (\ref{Y-dot-0}), hold for $H \ll m$, and so they do not hold exactly at the onset of the oscillations, when $H \lsim m$. This may introduce ${\cal O } \left( 1 \right)$ corrections in our derivation, but does not change the order of magnitude of our final result.}   $H=\frac{2}{3 t}$.  Eq. (\ref{eom}) then becomes a system of linear equations $\partial_0^2 \left( a^{3/2} \, \Phi_i \right) + \left( \partial_i \partial_j V \right) a^{3/2} \, \Phi_j \simeq 0$, which  can be diagonalized by solving the eigenvalue problem
\begin{equation}
\left( \partial_i \partial_j V \right) \, v_{n,j} = \lambda_n^2 \, v_{n,i} \,,\;\; v_{m,i} v_{n,i} = \delta_{mn} \,,
\label{eigenvalue}
\end{equation}
where we have introduced the eigenstates $Y_n \equiv~v_{n,i} \, a^{3/2} \, \Phi_{i}$ with eigenmasses $\lambda_n$. The evolution equation then becomes $\partial_0^2 Y_n = \lambda_n^2 \, Y_n$, and leads to the solution
\begin{equation}
Y_n \left( \tau \right) \simeq Y_n \left( t_0 \right) \cos \left( \lambda_n \, \tau \right) +  \frac{\dot{Y}_n \left( t_0 \right)}{\lambda_n} \sin \left( \lambda_n \, \tau \right) \,, 
\label{Y-sol}
\end{equation}
where $\tau = t-t_0$, and $t_0$ denotes the end of inflation. Setting $a \left( t_0 \right) = 1$, we have $Y_n \left( t_0 \right) = v_{n,j} \Phi_j \left( t_0 \right)$ and
\begin{equation}
\dot{Y}_n \left( t_0 \right) = v_{n,j} \left[ \dot{\Phi}_j \left( t_0 \right) + \frac{3}{2} H \left( t_0 \right) \Phi_j \left( t_0 \right) \right] \simeq v_{n,j} \, \dot{\Phi}_j \left( t_0 \right) \,,
\label{Y-dot-0}
\end{equation}
so that (\ref{Y-sol}) becomes
\begin{equation}
\Phi_i \left( \tau \right) \simeq \frac{v_{n,i} v_{n,j}}{a^{3/2}} \left[ \Phi_j \left( t_0 \right) \cos \left( \lambda_n \, \tau \right) +
\frac{\dot{\Phi}_j \left( t_0 \right)}{\lambda_n} \, \sin \left( \lambda_n \tau \right) \right].
\label{formal}
\end{equation}
Substituting this into (\ref{nL}) we obtain
\begin{eqnarray}
&& \!\!\!\! n_L =  C_{nilj} \Big\{ \left( \lambda_n -  \lambda_l \right) \sin \left[ \left( \lambda_n +   \lambda_l \right) \tau + \delta_{+,nilj} \right] \nonumber\\ 
&& \quad\quad\quad +  \left( \lambda_n +  \lambda_l \right) \sin \left[ \left( \lambda_n -   \lambda_l \right) \tau + \delta_{-,nilj} \right] \Big\} \,, 
\label{nL-formal}
\end{eqnarray}
where the constants are defined as
\begin{eqnarray}
&& \!\!\!\!\!\!\!\!
C_{nilj} \equiv \frac{1}{ 2a^3} \left(v_{n,1} v_{l,3}  + v_{n,2} v_{l,4}\right) v_{n,i} v_{l,j}  \nonumber\\
&& \quad\quad \times \, \sqrt{\Phi^2_i \left( t_0 \right) + \frac{\dot{\Phi}^2_i \left( t_0 \right)}{\lambda_n^2} }  \sqrt{\Phi^2_j \left( t_0 \right) + \frac{\dot{\Phi}_j^2 \left( t_0 \right)}{\lambda_l^2} } \,, \nonumber\\
&& \!\!\!\!\!\!\!\!
\delta_{\pm,nilj} \equiv \tan^{-1} \left[ \frac{\lambda_n \, \Phi_i \left( t_0 \right)}{\dot{\Phi}_i \left( t_0 \right)} \right] \pm  \, \tan^{-1} \left[ \frac{\lambda_l \, \Phi_j \left( t_0 \right)}{\dot{\Phi}_j \left( t_0 \right)} \right].
\end{eqnarray}
We see that the lepton-number-violating and $CP$-violating masses in $V$ cause the lepton number (\ref{nL-formal}) to oscillate with frequencies given by the sums and differences of the eigenmasses.

The eigenvalue problem (\ref{eigenvalue}) can be solved as an expansion series in  $\mu_{I} \ll 1$. We denote the mass-squared eigenvalues as $\lambda_{1,2}^2 \equiv\lambda^2_\pm$ and $\lambda_{3,4}^2\equiv\Lambda^2_\pm$, where
\begin{eqnarray}
&& \!\!\!\! \!\!\!\! \lambda^2_{\pm} \simeq m^2 \! \left[ \left( 4 + \mu_R^2 \right) P_\pm^2 + \mu_R \, P_\pm \, \mu_{3/2} \right] \!
+ \!  {\cal O } \left(  \mu_I^2 ,\, \mu_{3/2}^2 \right) \!, \nonumber\\
&& \!\!\!\! \!\!\!\!  \Lambda^2_\pm \simeq m^2 \! \left[ \left( 4 + \mu_R^2 \right) P_\pm^2 + \mu_R Q_\pm\mu_{3/2} \right] \!
+ \! {\cal O } \left(  \mu_I^2 ,\, \mu_{3/2}^2 \right) \!,
\label{eigenvalues}
\end{eqnarray}
and we have introduced the two parameters
\begin{equation}
P_\pm \equiv \frac{1}{2} \pm \frac{\mu_R}{2 \sqrt{4+\mu_R^2}} \;\;,\;\;
Q_\pm \equiv \frac{3}{2}\pm \frac{8 + 3 \mu_R^2}{2 \mu_R\sqrt{4+\mu_R^2}} \,.
\end{equation}

Substituting (\ref{eigenvalues}) and the expressions for the eigenvectors  into (\ref{nL-formal}), we find that only four terms proportional to $\sin \left[ \left( \lambda_\pm + \Lambda_\pm \right) \tau \right]$ and $\sin \left[ \left( \lambda_\pm - \Lambda_\pm \right) \tau \right]$ contribute to leading order in $\mu_I$. The  sums of the eigenmasses correspond to fast oscillating contributions to $n_L$, that  average away during the decay of the right-handed sneutrinos. The frequencies corresponding to the differences, are instead of ${\cal O } \left( m_{3/2} \right)$, and are therefore much smaller than the decay rate.\footnote{Ref. \cite{Allahverdi:2004ix} discussed an analogous mechanism for leptogenesis, but assumed that the right-handed sneutrino decays when the slow oscillating term is at a maximum. Here we make the more natural assumption of  a right-handed sneutrino decay rate $\Gamma \gg m_{3/2}$, resulting in a ${\cal O } \left( m_{3/2} \Gamma^{-1} \right)$ suppression of the asymmetry.} Neglecting the fast oscillating terms, we obtain
\begin{eqnarray}
\label{L}
&& \!\!\!\! \!\!\!\! n_L   \simeq   -\xi_0  \frac{ \mu_I \mu_{3/2} \, t \, m^2 \, M_p^2}{\mu_R^{2/3} a^3} + {\cal O } \left( \mu_I \mu_{3/2}^2 , \mu_I^2 \mu_{3/2} \right) \;, \\
&& \!\!\!\! \!\!\!\! \xi_0  \equiv  \frac{  \phi_{1I}(t_0)}{M_p }   \frac{\phi_{2I}(t_0)}{\mu_R^{1/3}M_p}     \!\! \left( \! 1 + \frac{\dot{\phi}_I^2(t_0)}{3 m^2 \phi_{1I}^2(t_0)} \! \right) +  {\cal O } \left( \mu_R^{2/3} \right)\,, \nonumber
\end{eqnarray}
in the time interval $m^{-1} \ll  t \ll m_{3/2}^{-1}$. The coefficient $\xi_0$  depends on the values of the fields at the end of inflation, which give $\xi_0 \simeq 1$.

\vspace{-3mm}
\section{Reheating and  Baryon Asymmetry}
\label{sec:reheating}
\vspace{-3mm}

To estimate the baryon asymmetry obtained from (\ref{L}), we assume an instantaneous decay of the sneutrino fields with the  rate $\Gamma \equiv \frac{h^2}{8 \pi} m$, where $h$ is the Yukawa coupling of the $N L H_u$ interaction ($L$ and $H_u$ denote the left-handed lepton and the up-type Higgs doublet, respectively). The Yukawa coupling is related to the neutrino masses
by the see-saw formula $m_\nu \simeq h^2 v_u^2/m$ (suppressing flavor indices) where $v_u =\langle H_u \rangle$. A fit to the neutrino oscillation data leads to Yukawa couplings as large as $h \sim 0.1$ for the largest 
 neutrino mass $m_\nu \sim 0.05$ eV. We stress however that the two neutrino fields employed in our mechanism do not necessarily need to be the two heaviest ones, so that $h$ can be smaller than $0.1$. 
 
 Before the decay, sneutrino oscillations dominate the energy density of the universe, enforcing the scale factor evolution $a= \left( m t \right)^{2/3}$. The  decay occurs when $H \left( t \right) = \Gamma$, and  generates a thermal bath with reheating temperature
\begin{equation}
T_{\rm RH} \simeq 4.4 \times 10^{13} \, {\rm GeV} \, \frac{h}{0.1} \, \sqrt{\frac{m}{10^{13} \, {\rm GeV}} } \,.
\label{TRH}
\end{equation}
The lepton number (\ref{L}) is transferred to the thermal bath, and is then partially reprocessed by electroweak sphalerons into the baryon number  $n_B=-\frac{8}{23} n_L$\,\cite{Giudice:2003jh}. This gives  the abundance
\begin{equation}
Y_B \equiv \frac{n_B}{s} \simeq \frac{\mu_I \, \mu_{3/2}}{27 \,\mu_R^{2/3}} \, \sqrt{\frac{M_p}{\Gamma}} \,,
\label{nB}
\end{equation}
where $s$ is the entropy density of the thermal bath \cite{Kolb:1990vq}. The $\frac{n_B}{s}$  ratio is unaffected by the expansion of the universe, and controls both the light-element abundance formed during big-bang nucleosynthesis and the height of the CMB peaks. The CMB constraint is the most stringent and enforces  $n_B/s \simeq 9\times 10^{-11}$ \cite{Ade:2013uln}. This can be satisfied by  (\ref{nB}), provided
\begin{equation}
 m_{3/2} \simeq 3.6 \, {\rm GeV} \,
 \frac{\mu_R}{\mu_I} \left( \frac{0.02}{\mu_R} \right)^{1/3} \frac{h}{0.1}\left( \frac{m}{10^{13} \, {\rm GeV}}\right)^{3/2}\,.
\label{m32-us}
\end{equation}

As in the thermal leptogenesis scenario  \cite{Fukugita:1990gb}, additional lepton asymmetry may be generated  at the decay of the right-handed sneutrino condensates, and of the neutrino and sneutrino quanta recreated by the thermal bath (notice that the reheating temperature (\ref{TRH}) is marginally greater than $m$). We assume that this contribution is negligible. This can be easily achieved if the CP-violating phases in the Yukawa coupling  are sufficiently small. Assuming negligible CP violation in $h$ also  ensures that the asymmetry (\ref{nB}) is not washed out by thermal scatterings between the lepton and Higgs quanta mediated by the right-handed neutrinos.

A high reheating temperature can also lead to a significant thermal gravitino production.  For $T_{RH}\simeq 10^{13}$~GeV the  gravitino abundance is $Y_{3/2} \simeq 2 \times 10^{-9}$~\cite{Kawasaki:2008qe}. Such gravitinos may be identified with the present dark matter if $m_{3/2} = {\cal O } \left( 100 \, {\rm MeV} \right)$. This is incompatible with (\ref{m32-us}), under the assumption that $\mu_I \ll \mu_R$. Instead, provided that $m_{3/2} \gtrsim 10 \, {\rm TeV}$
\cite{Kawasaki:2008qe},  we can  assume that gravitinos decay before nuclesynthesis. The decay produces the lightest supersymmetric particle (LSP) with an identical abundance to the gravitino parent, which can be identified with the current dark matter provided its mass is now of ${\cal O } \left( 100 \, {\rm MeV} \right)$ \cite{Dreiner:2009ic}. Another possibility is to consider R-parity violation, so that the LSP also decays \cite{Murayama:2014saa}, which, however, has the drawback that dark matter cannot be the LSP neutralino. A third possibility is to impose that gravitinos decay before the dark matter freezes out (namely, at a temperature $ \gtrsim  m_{\rm LSP} / 20$), so that the LSP produced by their decay thermalizes, and their final abundance is the thermal one (in this case, neutralino dark matter with electroweak mass can be assumed). This requires $m_{3/2} \gtrsim 10^7  {\rm GeV} \left(\frac{m_{\rm LSP}}{100\,{\rm GeV}}\right)^{2/3}$. This large 
 hierarchy between the gravitino and the LSP mass is a feature of certain models~\cite{Luty:2002ff,Gherghetta:2011wc}.

\vspace{-3mm}
\section{Conclusion}
\label{sec:conclusions}
\vspace{-3mm}

In this Letter we have presented a supergravity model where the inflaton is identified with a linear combination of right-handed sneutrino fields. Due to a shift symmetry the dominant component is absent from  the  K\"ahler potential. We have studied a minimal version of the model, characterized by two sneutrino fields and two mass terms. The resulting potential along the inflationary trajectory is flatter than that of quadratic chaotic inflation, giving a gravitational wave signal that can be detected in the near future. The normalization of the scalar perturbations fixes the right-handed  neutrino mass scale to the naturally expected value in the see-saw mechanism. CP violation in the neutrino masses and supersymmetry breaking cause a lepton charge carrying evolution of the sneutrino fields, that can be the origin of the observed baryon asymmetry of the universe. Thus, the smallness of the  asymmetry is naturally due to the double suppression from small CP violation and from a supersymmetry-breaking scale much below the Planck scale.

\vspace{0.5cm}
{\bf Acknowledgments}
This work is supported in part by the Department of Energy grant DE-SC0011842 at the University of Minnesota.

\end{document}